\documentclass[twocolumn,preprintnumbers,amsmath,amssymb,prl]{revtex4}
\usepackage{txfonts}
\usepackage{graphicx}
\usepackage{hyperref}
\usepackage{dcolumn}
\usepackage{bm}
\usepackage{doi}
\usepackage{textcomp}
\usepackage{setspace}
\usepackage{float}
\usepackage{indentfirst}
\usepackage{natbib}
\usepackage{subfigure}
\usepackage{booktabs}
\usepackage{multirow}
\usepackage{txfonts}
\usepackage{amsmath}
\usepackage{braket} 

\def\sech{\operatorname{sech}}

\def\exp{\operatorname{e}}
 \allowdisplaybreaks

\makeatletter

\newcommand{\Rmnum}[1]{\expandafter\@slowromancap\romannumeral #1@}
\makeatother

\hypersetup{colorlinks=true,allcolors=blue,urlcolor=cyan,pdfstartview=Fit,breaklinks=true}
\begin{document}
\bibliographystyle{plainnat}
\title{Controlled pumping of matter-wave solitons in a one-dimensional optical superlattice}
\author{Xiaoxiao Hu}
\affiliation{Department of Physics, Zhejiang Sci-Tech University, Hangzhou 310018, China}
\author{Zhiqiang Li}
\affiliation{Department of Physics, Zhejiang Sci-Tech University, Hangzhou 310018, China}
\author{Ai-Xi Chen}
\affiliation{Department of Physics, Zhejiang Sci-Tech University, Hangzhou 310018, China}
\author{Xiaobing Luo}
\email[]{Corresponding author: xiaobingluo2013@aliyun.com}
\affiliation{Department of Physics, Zhejiang Sci-Tech University, Hangzhou 310018, China}
\date{\today}
\begin{abstract}
We study the pumping of matter-wave solitons  formed in Bose-Einstein condensates (BECs) with attractive atomic interactions that are loaded into optical superlattices in which one of the lattices is moving with respect to the other. We find that solitons exhibit the remarkably similar pumping properties in both shallow and deep lattices, and that for exactly the same soliton initially excited, switching between integer (fractional) pumping and trapping can be achieved by simply adjusting the lattice parameters. In addition, we find that the gap solitons, which bifurcate from the lowest energy band in a semi-infinite band gap, also exhibit this lattice parameter dependent pumping and trapping. The treatment of solitons as classical particles with effective centre-of-mass equations of motion provides a good description of this parameter-dependent integer (fractional) pumping and trapping of solitons.
\end{abstract}
\maketitle
\paragraph{\textbf{Introduction:}}Quantum pumping is an important way to achieve directed transport of matter. A famous example is the Thouless pumping: Thouless~\cite{Thouless1983} showed that the one-dimensional (1D) pumping of charge by a slow-moving periodic potential over a pumping period is quantized, and that the amount of charge pumped is determined by the Chern number considered in the extended coordinate-time space. Thouless pumping has the same topological origin as the two-dimensional (2D) quantum Hall effect (QHE)~\cite{Klitzing1980} and is a topological protection of wave transport that is not limited to fermions. Thouless pumping is currently observed experimentally in cold bosonic~\cite{Lohse2016,Lohse2018} and fermionic~\cite{Nakajima2016,Nakajima2021,Walter2023} atoms, spin systems~\cite{Ma2018}, optical~\cite{Kraus2012,Zilberberg2018,Cerjan2020}, acoustic~\cite{Cheng2020} and plasma~\cite{Fedorova2020} systems.

Recently, the exploration of topological quantum transport in nonlinear and interacting systems has attracted great interest. It has been exciting to observe trapped solitons, and quantized integer and fractional Thouless pumping~\cite{M2021,M2023} of solitons in nonlinear optical waveguide arrays (see recent reviews~\cite{Citro2023}). Despite non-uniform linear band occupation, the motion of low-power soliton shows exactly an integer-quantized nonlinear Thouless pumping~\cite{M2021}, which is dictated by the Chern number of the band from which the soliton bifurcates~\cite{M2022,Mostaan2022}. As the nonlinearity (soliton power) increases, the soliton starts to occupy energy eigenstates of higher bands and undergoes the transition from integer-quantized pumping to a fractionally quantized pumping and finally to trapping. It has been theoretically analyzed that the strongly nonlinear soliton follows the position of the instantaneous maximally localized multi-band Wannier functions~\cite{M2023}, and that the centre-of-mass displacement of the fractional soliton pumping is related to the chern numbers of the participating linear bands~\cite{Fu2022,Ye2022}.

In this letter, we address the nonlinear pumping of matter-wave solitons formed in a BEC with attractive atomic interactions loaded in a slowly varying superlattice potential. We find that whether in shallow or deep lattices, solitons exhibit the remarkably similar pumping characteristics, and that for the initial excitation of the same soliton, the transition between integer (fractional) pumping and trapping can be achieved by adjusting the lattice parameters. The nonlinear soliton pump can be understood by treating solitons as effective classical particles. Of particular importance is that this transition from a trapped soliton to a pumped soliton with increasing nonlinearity is, to the best of our knowledge, found for the first time in a nonlinear soliton pump. In addition, we find that gap solitons that bifurcate from the lowest energy band  in a semi-infinite band gap also exhibit this lattice-parameter-dependent pumping and trapping, and their relationship to the Chern number is investigated. Different from the perspective from the linear band Chern numbers, we find that the classical particle properties of solitons in BEC play a crucial role in soliton pumping.

\paragraph{\textbf{Model:}}We consider a matter-wave soliton formed in a highly elongated condensate cloud subjected to a dynamical 1D optical superlattice, which consists of a stationary lattice created by two counter-propagating monochromatic laser beams~\cite{Peil2003,Morsch2006,Zenesini2010} and a sliding lattice created by two counter-propagating beams  that are detuned in frequency~\cite{Lignier2007,Atala2013}. As long as the energy of the longitudinal excitations is not sufficient to excite the transverse modes of the BEC, the system can be described by the 1D Gross-Pitaevskii equation(GPE):
\begin{equation}\label{con:1}
i \Psi_{{t}}=-\frac{1}{2} \Psi_{{x} {x}}-|\Psi|^{2} \Psi+V(x,t)\Psi.
\end{equation}
Here $t$ and $x$ are measured in units of $\frac{\hbar}{2E_r}$ and $\frac{\Lambda}{\pi}$ respectively, $E_r=\frac{\hbar^2\pi^2}{2m\Lambda^2}$ is the recoil energy, $m$ is the atomic mass, and $\Lambda$ is a characteristic length defining the period of the potential $V(x,t)$ whose amplitude is measured in units of $2E_r$. For a BEC with negative scattering length ($a_s<0$) in a cigar-shaped trap with transversal trap frequency $\omega_{\perp}$, the dimensionless $\Psi$ and the dimensional $\Phi$ order parameter are related by $\Phi=\sqrt{\hbar\omega_{\perp}|a_s|/E_r}\Psi$ (see, e.g., ~\cite{V1998}). The dynamical superlattice is of the form~\cite{Lohse2016}
\begin{equation}\label{con:2}
V(x,t)=-p_{1} \cos ^{2}\left(\pi x / d_{1}\right)-p_{2} \cos ^{2}\left(\pi x / d_{2}-\nu t\right),
\end{equation}
where $p_{1,2}$ and $d_{1,2}$ are the dimensionless depths (in units of $2E_r$) and the periods of the constitutive lattices. The periods $d_1$ and $d_2$ are considered to be commensurate, i.e. $d_1/d_2=n_1/n_2$, where $n_1$ and $n_2$ are coprime integers. Here, the relative position of the lattices is determined by the variable phase $-\nu t$, where we require $\nu$ to be a small parameter that determines the adiabatic displacement of the second lattice. The lattice then changes periodically in time with period $T=\pi/\nu$. At each time instant the potential remains periodic with the dimensionless period $L= n_1d_2 = n_2d_1$. In the linear case, the system containing the external potential of Eq.\;\eqref{con:2} has been known to exhibit quantized pumping~\cite{Lohse2016,Nakajima2016,Wang2013}, also known as Thouless pumping. For the nonlinear evolution, we consider the experimental setup of Refs.~\cite{Khaykovich2002,Strecker2002}, where a bright soliton forming in the $^{7}$Li cloud with a modified scattering length $a_s=-1. 43$ $\mathrm{nm}$ is trapped in a quasi-one-dimensional atomic waveguide with $\omega_{\perp}=2\pi\times710$ $\mathrm{Hz}$ and $\Lambda=2$ \textmu m. Given these physical parameters, the time period $T=10\pi$ and the distance $L=1$ used in our simulations are equivalent to $4.2$ $\mathrm{ms}$ and $0.64$ \textmu m, respectively, and a
stable bright soliton typically created in the experiment contains $10^3$ atoms, which corresponds to $N=\int_{-\infty}^{\infty}|\Psi({x})|^{2} d {x}\approx8.5$.

In the absence of the dynamical optical superlattice, the Eq.\;\eqref{con:1}  has a well-known solution in the shape of a moving bright soliton:
\begin{equation}\label{con:3}
\Psi(x,t)=\frac{N}{2}\sech \left[\frac{N}{2}(x-x_0)\right]\exp^{i[v_0(x-x_0)-\mu t]},
\end{equation}
where $x_0$ is the position of the center of mass of the soliton, $v_0$ is its velocity and $\mu$ is the chemical potential. It can be seen from Eq.\;\eqref{con:3} that the soliton is strongly localized for large $N$, while it is more extended for small $N$. We first consider the case where the shape of the solitons is not significantly changed during the pump, which is possible in the system with relatively shallow lattices.

\paragraph{\textbf{Shallow superlattice:}}In the general case, the GP system\;\eqref{con:1} is not integrable. We therefore use a variational approximation to investigate the soliton dynamics~\cite{Malomed2006,Malomed2002}. Using the moving soliton solution $\Psi(x,t)$, the Lagrangian corresponding to Eq.\;\eqref{con:1}, namely, $\mathcal{L}(t)=\int_{-\infty}^{\infty}\left\{\frac{i}{2}[\Psi^\ast\Psi_t-\Psi(\Psi_t)^\ast]-(\frac{1}{2}|\Psi_x|^2-\frac{1}{2}|\Psi|^4+V|\Psi|^2)\right\}dx$, gives the effective equation (EE) of motion for the soliton's center of mass
\begin{equation}\label{con:4}
 \ddot{{x}}_0(t)=-\frac{2\pi^3}{N}\left[\frac{p_1\sin(\frac{2\pi}{d_1}x_0)}{d_1^2\sinh(\frac{2\pi^2}{Nd_1})}+\frac{p_2\sin(\frac{2\pi}{d_2}x_0-2vt)}{d_2^2\sinh(\frac{2\pi^2}{Nd_2})}\right],
\end{equation}
which describes the matter-wave soliton as a classical particle moving in an effective potential.

\begin{figure}[htbp]
	\centering
\includegraphics[trim=20 0 0 0,clip,width=9.2cm]{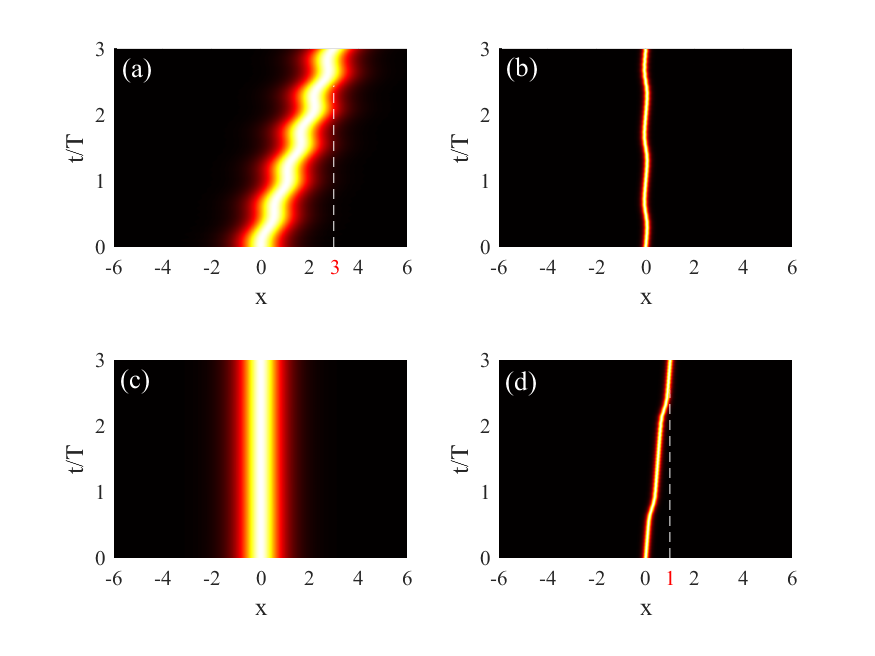}
    \caption{Evolution of the initial state $\Psi(x,0)$ with $v_0=0$ (i.e. a bright soliton) in shallow lattices for different normalized particle numbers. (a) and (c) $N=4$; (b) and (d) $N=30$. In panels (a) and (b), $d_2 = 1>d_1$, the transition from pumping to trapping with increasing $N$ is observed, whereas in panels (c) and (d), $d_2 = 1/3<d_1$, the transition from trapping to pumping  with increasing $N$ is observed. In all cases, $d_1=\frac{1}{2}$, $p_1=p_2=0.1$, $v=0.1$, $T=10\pi$.}
    \label{fig1}
\end{figure}

The effective equation\;\eqref{con:4} shows that the motion of the center of mass of the soliton is influenced by two main parts (see the Supplemental Material~\cite{Material}). The first part comes from the stationary optical lattice, which corresponds to an effective potential well that leads to soliton trapping. The second part comes from a sliding optical lattice, where the slow and uniform sliding of the lattice is the cause of the pumping of the solitons. For small $N$, when $d_2 > d_1$, the first term on the right-hand side of Eq.\;\eqref{con:4} is exponentially smaller than the second term, then the pumping dynamics is dominated by the sliding potential; when $d_2 < d_1$, the first term dominates and the effective potential becomes virtually independent of the second term (the sliding potential). The opposite is true for large $N$. From these considerations, we expect the nonlinear pumping dynamics to depend on $N$ and the relative magnitudes of $d_1$ and $d_2$. In Fig.\;\ref{fig1}, we show the soliton evolution for small particle numbers [see panels (a) and (c)] and large particle numbers [see panels (b) and (d)] calculated for three periods by direct simulation of the GP equation\;\eqref{con:1} with the initial excitation given by Eq.\;\eqref{con:3} with $v_0=0$ at $t = 0$ (that is  a static soliton). As shown in Figs.\;\ref{fig1}(a)-(b), for the case $d_2>d_1$, the soliton with small $N$ is pumped to the left (in the positive direction) by three unit cells after three periods, while the soliton with large $N$ is trapped. The situation is reversed for the case $d_2<d_1$, where the soliton with small $N$ becomes trapped while the fractionally pumped but not integer-pumped soliton (the center of mass displacement is $1/3$ for one period) is observed with large $N$, as shown in Figs.\;\ref{fig1}(c)-(d).

\begin{figure}[htbp]
	\centering
\includegraphics[trim=22 0 0 0,clip,width=9.2cm]{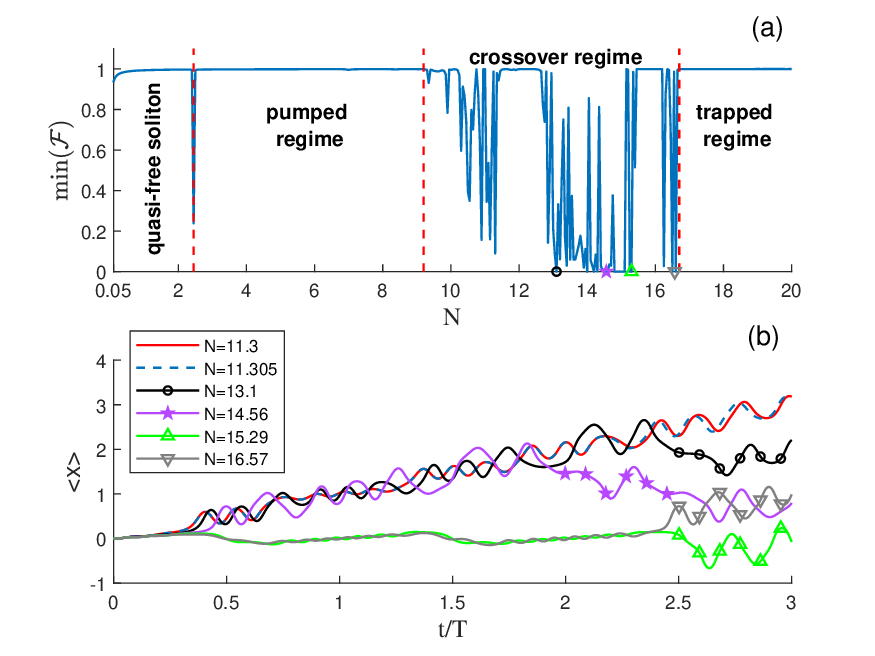}
    \caption{Dynamical stability of soliton evolution for different regimes in shallow lattices. (a) The minimum value of the fidelity given by Eq.\;\eqref{con:5} versus $N$ for the soliton evolution. (b) The unstable transports of solitons in the crossover regime. $p_1=p_2=0.1$, $v=0.1$, $T=10\pi$, $\phi=0$, $d_1=\frac{1}{2}, d_2=1$. The initial excitation is given by Eq.\;\eqref{con:3} with $v_0=0$.}
    \label{fig2}
\end{figure}

As shown above, the nature of the soliton evolution can be varied by changing the atomic numbers, the lattice constant or the pump potential. To explain the nonlinear transition between soliton pumping and trapping, we examine separately the single sliding sublattice and the single static sublattice, which are shown to be the principle causes of the soliton pumping and trapping, respectively. For a shallow sliding lattice ($p_1=0$), the system can be mapped to a classical pendulum by introducing the pendulum frequency $\omega_0=\sqrt{\frac{4\pi^4p_2}{Nd_2^3\sinh(\frac{2\pi^2}{Nd_2})}}$ (see Supplemental Material~\cite{Material}). In the pendulum analogy, the main properties of soliton pumping in a single sliding lattice are demonstrated analytically: For a sufficiently slow sliding lattice ($\nu<\omega_0$) we can obtain a classical adiabatic pumping where the soliton follows the sliding lattice, pumping by a distance of about $d_2$ (either integer or fractional) per cycle. For a stationary  shallow lattice ($p_2= 0$), the stationary optical lattice is equivalent to a potential well~\cite{Scharf1993}, the depth of which increases with the particle number of the soliton (see Supplemental Material~\cite{Material} for details). This suggests that if the pumping term is not sufficient for the soliton to escape from the effective potential well, it will be captured by the effective potential well, resulting in soliton trapping. Thus, whether the soliton is trapped or pumped depends on the competition between the pumping term (caused by the sliding lattice) and the trapping term (caused by the stationary lattice), and thus on the size of $N$, as well as the relative sizes of $d_1$ and $d_2$. In numerical simulations, we probe three stable regimes (quasi-free, pumped, trapped) when scanning the particle numbers. Interestingly, the soliton exhibits a nonlinear transition regime between pumping and trapping, which is connected with the instability of the mean-field dynamics. In Fig.\;\ref{fig2}(a), we use the minimum value of quantum fidelity $\mathcal{F}$ achieved during the time evolution to quantify the divergence of nearby trajectories. The quantum fidelity $\mathcal{F}$ is conventionally defined as
\begin{equation}\label{con:5}
\mathcal{F}(t)=\left|\int_{-\infty}^{\infty}dx\Psi^{*}(x,t)\tilde{\Psi}(x,t)\right|^2/N^2,
\end{equation}
where the denominator $N^2$ is introduced for normalization. and $\Psi(x,t)$ and $\tilde{\Psi}(x,t)$ are two time-evolving states starting from $\Psi(x,0)$ and its slightly perturbed state $\tilde{\Psi}(x,0)$. Clearly, if $\mathrm{min}(\mathcal{F})$ is close to $1$, the state is dynamically stable, otherwise it is dynamically unstable.

As shown in Fig.\;\ref{fig2}(a), there exists a crossover regime (which is linked to the dynamical instability) from the stable pumped to the stable trapped solitons, with the change of the normalized particle number $N$. In Fig.\;\ref{fig2}(b), in the crossover regime between pumping and trapping, we numerically demonstrate the chaotic nature (dynamical instability) of the pumping transport and the transition from the pumping solution to the localized solution with time (see Supplemental Material~\cite{Material} for details). When $N < 2.45$, which is what we call the quasi-free regime, the size of the soliton is comparable to, or larger than, one spatial cycle of the dynamical superlattice. It is therefore more accurately described as a wave packet than as an effective particle. In this case, the right-hand side terms of Eq.\;\eqref{con:4} are both so small (due to the very small $N$) that the soliton hardly feels the effective superlattice and hardly moves if it is initially at rest. The instability of the soliton dynamics at $N=2.45$ is indicative of the transition from the wave packet to the particle nature of the soliton. So far we have only focused on the transition for the case $d_2>d_1$, as shown in Fig.\;\ref{fig2}. The approach and results will be very similar if instead we consider the case $d_2<d_1$, where the crossover regime with dynamical instability from the trapped soliton to a pumped soliton is observed as the normalized particle number $N$ is increased.

\begin{figure}[htbp]
	\centering
\includegraphics[trim=25 0 0 0,clip,width=9.2cm]{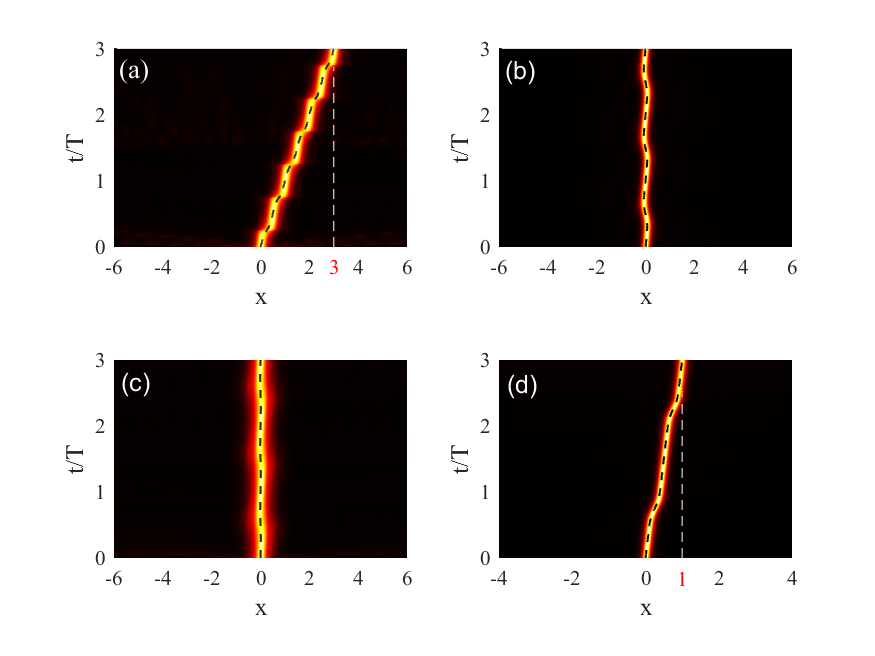}
    \caption{Pumping and trapping of solitons in deep optical lattices. (a), (c) $N=7$, (b) $N=20$, and (d) $N=30$. The transition from pumped soliton to trapped soliton with increasing $N$ is reversed to the transition from trapped soliton to pumped soliton by tuning $d_2=1>d_1$ [(a) and (b)] to $d_2=\frac{1}{3}<d_1$ [(c) and (d)]. The initial excitations are given by Eq.\;\eqref{con:3} with $v_0=0$. In all panels, the black dashed line visualizes the center-of-mass trajectory predicted by effective-particle dynamics\;\eqref{con:4}. The other parameters are $d_1=\frac{1}{2}$, $p_1=p_2=15$, $v=0.1$, $T=10\pi$.}
    \label{fig3}
\end{figure}

\paragraph{\textbf{Deep superlattice:}}For deep lattices, we can obtain a perturbed GP equation by introducing the renormalized order parameter \cite{Fu2022}. Here we consider the renormalized order parameter $\psi=\exp^{-i(p_1+p_2)\frac{t}{2}}\Psi/N$, and the equation for $\psi$ takes the form of the GP equation as follows
\begin{equation}\label{con:6}
i \psi_{{\tau}}+\frac{1}{2} \psi_{{X} {X}}+|\psi|^{2} \psi=\tilde{V}(X,\tau)\psi,
\end{equation}
where we have introduced the scaled variables $\tau=N^2t$ and $X=Nx$, and potential $\tilde{V}=-\frac{p_1}{2N^2}\left[\cos(k_1X)+p\cos(k_2X-\omega \tau)\right]$, with $p=p_2/p_1$, $\omega=2v/N^2$, and $k_j=2\pi/(Nd_j)$ ($j=1,2$). As long as $\frac{p_j}{2N^2}\ll1$, the moving soliton $\psi(X,\tau)=\frac{1}{2}\sech \left[\frac{1}{2}(X-X_0)\right]\exp^{iv_0(\tau)(X-X_0)}$, can be regarded as perturbed by $\tilde{V}(X,\tau)$ and remains essentially unchanged in shape during its motion. For sufficiently large $N$ it is ensured that $\frac{p_j}{2N^2}\ll1$, suggesting that our investigations into the nature of soliton pumping for shallow lattices may also be applicable to deep lattices. As shown in Fig.\;\ref{fig3}, for a deep superlattice, numerical results indicate that the soliton dynamics also follow the predictions of the effective equation\;\eqref{con:4}, see the black dashed lines. For solitons in the transition region between stable pumping and stable trapping, the stationary optical lattice induces high-frequency oscillations in the centre-of-mass displacement during pumping. This can lead to collective excitation of solitons and radiation generation (see Supplemental Material~\cite{Material} for details).

In particular, pumping and trapping of strongly nonlinear solitons, which can be characterized by the underlying linear topology, have been observed in the off-diagonal Aubry-Andr$\acute{\mathrm{e}}$-Harper model of optical system design~\cite{M2021,M2023}. Nevertheless, as shown in Fig.\;\ref{fig3}, in our model, the trapping and pumping of the soliton can be switched by adjusting $d_1$ and $d_2$, and for solitons with intermediate-mass (e.g. $N=7$), both trapped and pumped solitons can be observed, depending on the relative values of $d_1$ and $d_2$. When soliton pumping occurs, the soliton is displaced by a distance of $d_2$ after one pump cycle. By representing the soliton solutions in terms of linear Wannier orbitals, we find that the intermediate-mass soliton solutions ($N=7$ in Fig.\;\ref{fig3}) predominantly occupy the first lowest energy band with the well-defined Chern number (see Supplemental material ~\cite{Material} for more details), but the solitons exhibit trapping or pumping depending on the magnitudes of $d_1$ and $d_2$. This suggests that nonlinear solitons pumping presented here is an important consequence of their particle nature in dynamical optical lattices, which is not fully determined by the ratio of the populations among the involved linear bands.

\paragraph{\textbf{gap solitons:}}It is well known that nonlinear periodic systems allow for the existence of spatially localized eigenstates, the so-called gap solitons. When the gap solitons (nonlinear eigenstates) are initially excited, it is natural to ask whether  there is an equally nonlinear transition between pumping and trapping, and whether this transition can be tuned by lattice parameters and particle number, as is the case for the bright solitons in the form of Eq.\;\eqref{con:3} discussed above. We will give an affirmative answer based on numerical analysis.
\begin{figure}[htbp]
	\centering
\includegraphics[trim=28 0 0 0,clip,width=9cm]{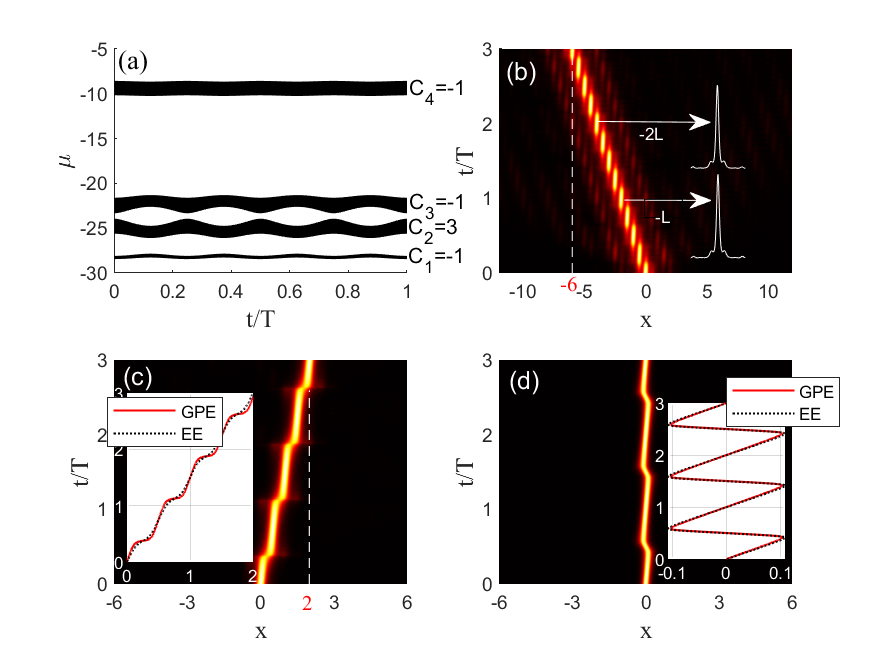}
    \caption{(a) The one-cycle evolution of the four lowest linear bands and their Chern numbers. (b) The quantized integer Thouless pumping at weak nonlinearity. The inset shows the shape of the solitons after one and two periods with the displacements in units of the lattice vector $L$. (c) The fractional pumping.  (d) The trapped soliton. The insets in (c) and (d) show the time evolution of the center-of-mass displacement obtained from the GP equation\;\eqref{con:1} (solid lines) and the effective equation\;\eqref{con:4} (dotted lines). The excitation is an instantaneous nonlinear eigenstate (that is, a gap soliton) of the system with different  degrees of nonlinearity (normalized particle number) (b) $N=0.3$, (c) $N=7$, and (d) $N=20$. Other parameters are $d_1=\frac{1}{2},d_2=\frac{2}{3}$, $p_1=p_2=25$, $v=0.1$, $T=10\pi$.}
    \label{fig4}
\end{figure}

Fig.\;\ref{fig4} shows the pump of the gap solitons that bifurcate from the lower band with Chern number $C_1=-1$. The band structure (instantaneous energy eigenvalues) of the Hamiltonian is shown in Fig.\;\ref{fig4}(a), where the four lowest bands (grey) with Chern numbers $C = [-1,3,-1,-1]$ are presented. As revealed in previous work~\cite{M2022,Mostaan2022}, at low nonlinearities, the gap soliton returns to its initial state after one pumping cycle, but shifts by one unit cell (lattice vector $L = 2$) in the negative direction (to the left), as shown in Fig.\;\ref{fig4}(b), corresponding to the Chern number of the lowest band. As the nonlinearity increases (the corresponding normalized particle number is increased to $N=7$), the soliton motion exhibits fractional pumping, where the soliton is pumped by $2/3$ per cycle in the positive direction, as seen in Fig.\;\ref{fig4}(c). Surprisingly, although the soliton shape changes during adiabatic evolution, the fractional center-of-mass displacement of $2/3$ can be well captured by the effective-particle dynamics\;\eqref{con:4}, as shown in the inset of Fig.\;\ref{fig4}(c), where the one-cycle-averaged displacement is governed by the lattice parameter $d_2$ (see Supplemental material~\cite{Material} for more details). This suggests that the fractional quantization of soliton pumping in our model may be accidental and not topologically protected by quantum mechanics~\cite{Wang2013}. For stronger nonlinearities ($N=20$ in our simulations) the soliton becomes trapped. It should be noted that, as in the case of the bright soliton discussed earlier, there is also a crossover regime between the stable, accidentally quantized pumping and the stable trapping, where a significant amount of radiation is emitted when the gap soliton is pumped.

So far we have only considered the pumping of the gap soliton for the case $d_2 > d_1$ as shown in Fig.\;\ref{fig4}, where the fractionally pumped soliton and a transition to a trapped soliton are clearly observed as $N$ is increased. Our numerical simulation shows that the motion of the gap soliton for the case $d_2 < d_1$ can also be predicted by the adiabatic effective-particle approximation, where the transition direction is reversed compared to the case $d_2 > d_1$, i.e. the transition from trapped to pumped soliton takes place for increasing $N$. In addition, we find that the closing of the bandgap between the two lowest bands (and thus having no well-defined Chern number) does not alter the transition between fractional pumping and trapping of solitons (see Supplemental material~\cite{Material} for more details). This further reveals the inspiring insight: the classical particle nature of the solitons plays a key role in the soliton pumping.

\paragraph{\textbf{Conclusion:}}In conclusion, we explain the mechanism of soliton pumping and trapping in dynamical superlattices and find that the high degree of nonlinearity will lead to the accidentally quantized pumped soliton. If the soliton is pumped, the one-cycle center-of-mass displacement is determined by the lattice parameter $d_2$ (whether integer or fractional). Our results are expected to have applications in cold atomic or optical systems, applicable to both shallow and deep lattices where the tight-binding model is effective.

The work was supported by the National Natural Science Foundation of China (Grant No. 12375022), and the Zhejiang Sci-Tech University Scientific Research Start-up Fund (Grant No. 20062318-Y).

\appendix
\section{Supplementary Material}
\subsection{Pumping of solitons in a single shallow sliding lattice }
In this section we discuss analytically the pumping of solitons in a single shallow sliding lattice. Considering the case $p_1 = 0$ from Eq.\;\eqref{con:4} in the main text, we obtain the effective centre-of-mass equation of motion for solitons in a sliding lattice
\begin{equation}\label{con:S1}
\ddot{{x_0}}(t)=-\frac{2\pi^3p_2}{Nd_2^2\sinh(\frac{2\pi^2}{Nd_2})}\sin(\frac{2\pi}{d_2}x_0-2v t).
\end{equation}
In the reference frame of the sliding lattice, by defining a new variable $\theta=2\pi(\frac{x_0}{d_2}-\frac{t}{T})$, where $T=\pi/v$, and by introducing $\omega_0=\sqrt{\frac{4\pi^4p_2}{Nd_2^3\sinh(\frac{2\pi^2}{Nd_2})}}$, Eq.\;\eqref{con:S1} can be written as a simple pendulum equation
\begin{equation}\label{con:S2}
\ddot{\theta}+\omega_0^2\sin(\theta)=0.
\end{equation}

To solve Eq.\;\eqref{con:S2}, different initial conditions determine different physical trajectories. Eq.\;\eqref{con:S2} can be solved exactly as
\begin{equation}\label{con:S3}
\theta(t)=\left\{
\begin{array}{cl}
& 2\arcsin\left\{\mathrm{sn}\left[F(\sin\frac{\theta(0)}{2},\frac{1}{k})\pm\omega_0 kt,\frac{1}{k} \right]\right\}, k \geq 1 \\
& 2\arcsin\left\{k \mathrm{sn}\left[F(k^{-1}\sin\frac{\theta(0)}{2},k)\pm\omega_0 t,k \right]\right\},  0 \leq k < 1 \\
\end{array} \right.
\end{equation}
where $F(y,k)=\int_{0}^{y}\frac{dy}{\sqrt{(1-y^2)(1-k^2y^2)}}$ is the Legendre elliptic integral of the first kind with $k$ being the modulus. The Jacobian elliptic function $\mathrm{sn}(u,k)$ is a function of the variable $u$ that oscillates in a sine-like manner between $+1$ and $-1$, with period $4K(k)$, where $K(k)\equiv F(1,k)$ denotes a complete elliptic integral of the first kind. The sign $\verb+"+\pm\verb+"+$ in Eq.\;\eqref{con:S3} is the same as that of the initial angular velocity $\dot{\theta}(0)$ of the pendulum, and the elliptic modulus $k =\sqrt{\sin^2\frac{\theta(0)}{2}+\frac{\dot{\theta}(0)}{4\omega_0^2}}$ is related to the initial position and the initial angular velocity of the pendulum.
\begin{figure}[htbp]
	\centering
\includegraphics[trim=0 0 0 0,clip,width=8cm]{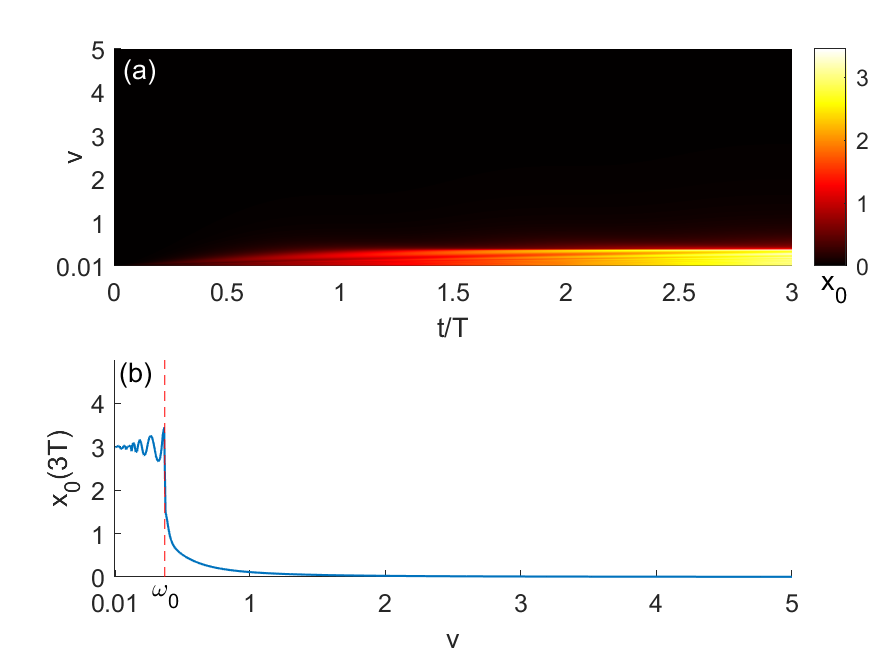}
    \caption{Centre-of-mass motion of the bright solitons in the single shallow sliding lattices, given by the numerical solutions of Eq.\;\eqref{con:S1} with the initial condition $\theta_0=0, v_0=0$. (a) Time evolution of the soliton position for different sliding lattice velocities. (b) The center-of-mass displacement of the soliton after three full periods versus the sliding lattice velocity. Other parameters chosen are $N=4, p_2=0.1, d_2=1$.}
    \label{figS1}
\end{figure}

The change in the position of the soliton centre of mass in the laboratory frame of reference is $x_0=d_2\frac{t}{T}+d_2\frac{\theta(t)}{2\pi}$, which contains a linear drift term $t/T$, which is naturally quantized at multiple cycles, and an oscillatory term $\theta(t)/2\pi$. The oscillation of $\theta(t)$ is given by Eq.\;\eqref{con:S3}, which of course depends on the initial conditions. We consider a simple case where the initial condition is $\theta(0)=0$ and $\dot{\theta}(0)=-\frac{2\pi}{T}$, such that $k =\frac{\pi}{\omega_0 T}$. In the limit case $T\longrightarrow0$, i.e. $k\longrightarrow+\infty$, there exists $\theta(t)\longrightarrow-\frac{2\pi t}{T}$, and such that the linear drift term is cancelled out, leading to $x_0\longrightarrow0$ and the trapping of the soliton. This corresponds to a fast pumping process in which the soliton is unable to follow the sliding lattice. In the opposite case, $T\longrightarrow\infty$, i.e. $k\longrightarrow0$, there is $\theta(t)\longrightarrow0$ and $x_ 0\longrightarrow d_2\frac{t}{T}$, so that the soliton is pumped by a distance close to $d_2$ after one cycle, which corresponds to an adiabatic pumping where the soliton follows the sliding lattice. The analysis shows that there are two different situations for the center-of-mass displacements of the soliton, corresponding to two different solutions of $\theta(t)$ given by  Eq.\;\eqref{con:S3} dependent on the value of the elliptic modulus $k$. As shown in Fig.\;\ref{figS1}, the two different states of motion of the soliton  can be distinguished by the point $k=1$ (corresponding to $v=w_ 0$) where the two solutions of $\theta(t)$ are connected. When $k>1$, i.e. $v>w_ 0$, there exists a fast pumping  such that the soliton can not follow the sliding lattice. When $k<1$, i.e., $v<w_ 0$, we obtain an adiabatic pumping, and the maximum amplitude of $\theta$ in the oscillatory term is $\theta_{max}=2\arcsin(\frac{\pi}{\omega_0 T})$. For slow pumping, we can expand $\theta_{max}$ to leading order, that is, $\theta_{max}\approx\frac{2\pi}{\omega_0 T}$, which means that the deviation from exact quantization is inversely proportional to $T$. This is in contrast to quantization in adiabatic quantum transport, which is protected by the energy gap and becomes exponentially accurate as $T$ increases~\cite{Wang2013}.

\subsection{Trapping solitons in a single static shallow lattice}
\begin{figure}[htbp]
	\centering
\includegraphics[width=7cm]{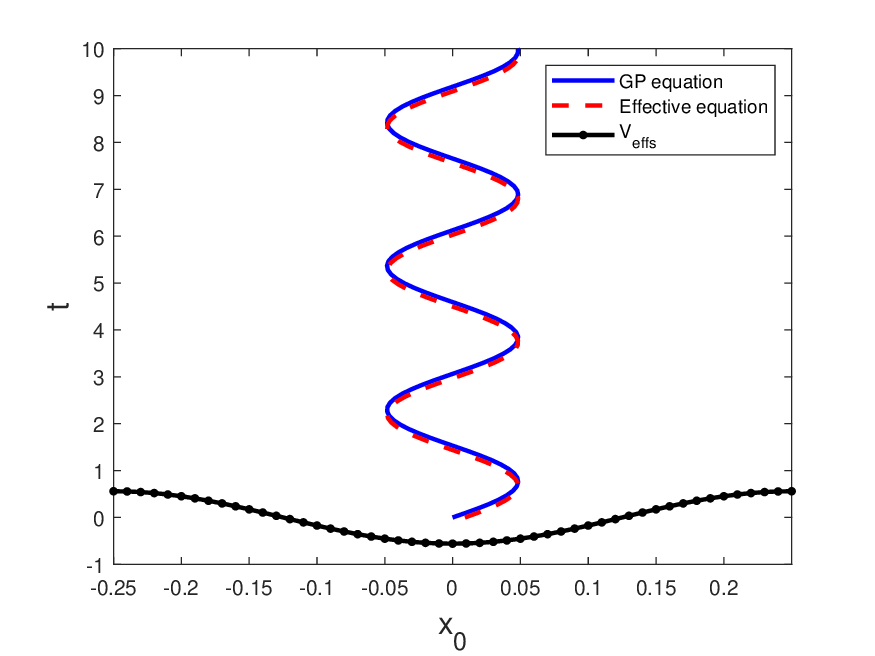}
    \caption{The effective potential $V_{\mathrm{effs}}$ given in Eq.\;\eqref{con:S5}, for the soliton moving the single static lattice. The change of the position of the soliton (blue solid line) predicted by the GP equation is compared with that of the effective-particle dynamics\;\eqref{con:S4} (red dashed line).}
    \label{figS2}
\end{figure}
In this section, we mainly discuss the role of the trapping of solitons by the single stationary lattice. In the case of a stationary lattice ($p_2= 0$), the effective equation of motion for soliton position [see the equation\;\eqref{con:4} in the main text] becomes
\begin{equation}\label{con:S4}
 \ddot{{x_0}}(t)=-\frac{2\pi^3p_1}{Nd_1^2\sinh(\frac{2\pi^2}{Nd_2})}\sin(\frac{2\pi}{d_1}x_0).
\end{equation}
The effective potential of Eq.\;\eqref{con:S4} is
\begin{equation}\label{con:S5}
V_{\mathrm{effs}}(x_0)=-\pi^2\frac{p_1\cos(\frac{2\pi}{d_1}x_0)}{d_1\sinh(\frac{2\pi^2}{Nd_1})}.
\end{equation}

For a soliton moving in the stationary optical lattice, the stationary lattice acts mainly as a confinement and the depth of the effective potential well increases with $N$. Fig.\;\ref{figS2} shows an example of soliton motion in a stationary lattice. The soliton motion obtained directly from GP equation is compared with predictions from effective particle dynamics and there is very good agreement between them. We can see that the soliton with an initial velocity of $v_0=0.1$ is trapped by the stationary optical lattice.
\subsection{Chaotic pumping of solitons in shallow dynamical superlattices}
\begin{figure}[htbp]
	\centering
\includegraphics[width=4cm]{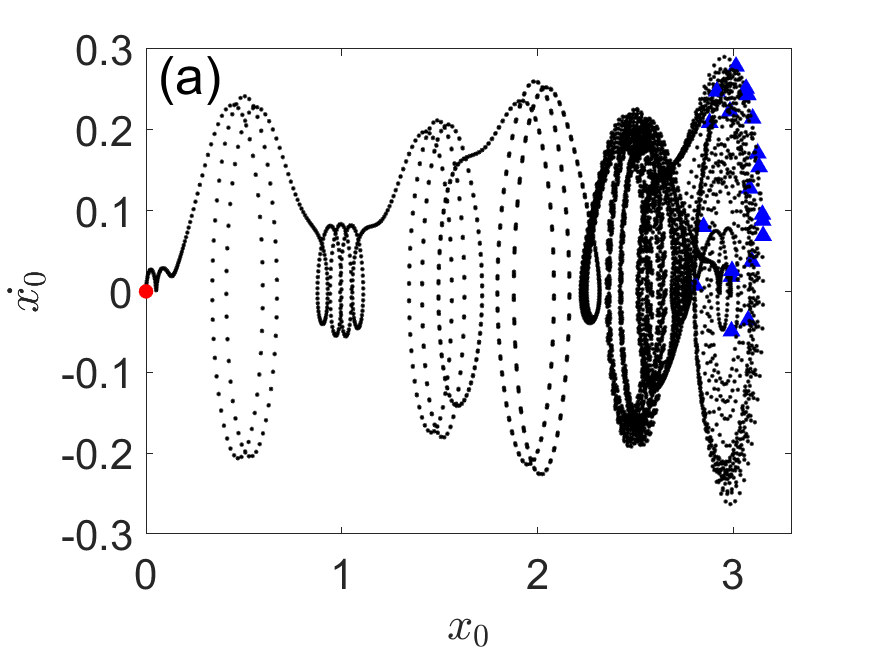}
\includegraphics[width=4cm]{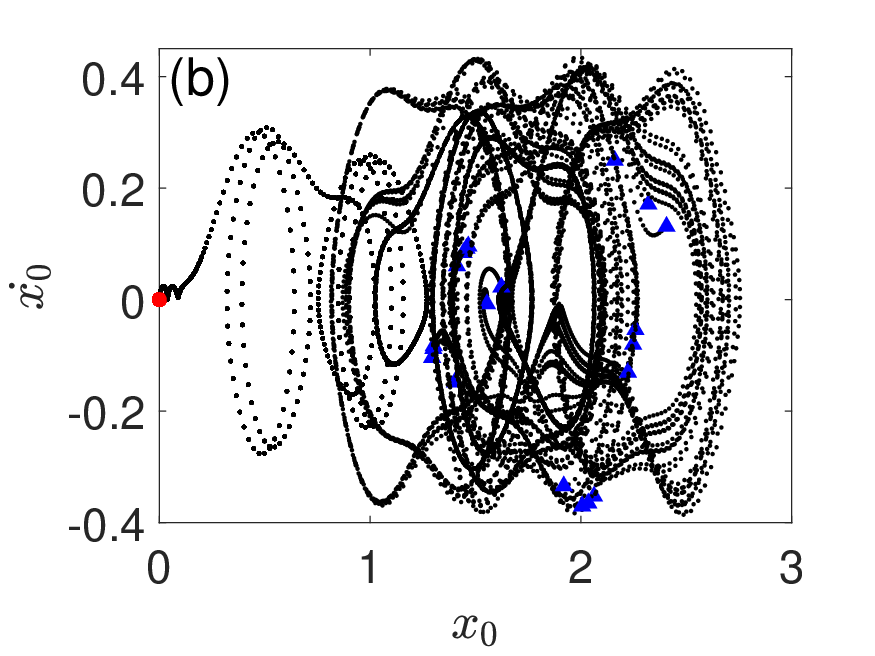}
\includegraphics[width=4cm]{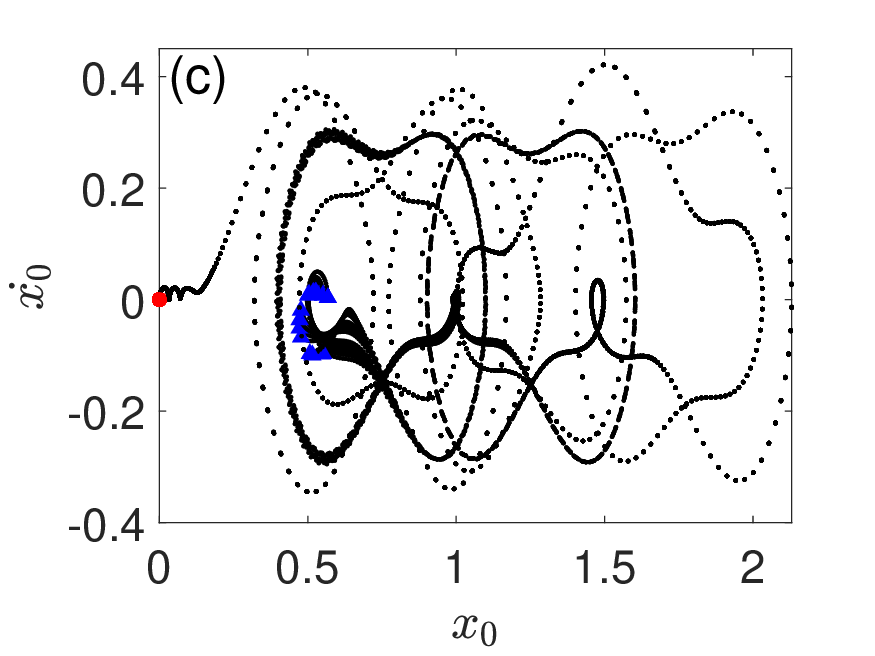}
\includegraphics[width=4cm]{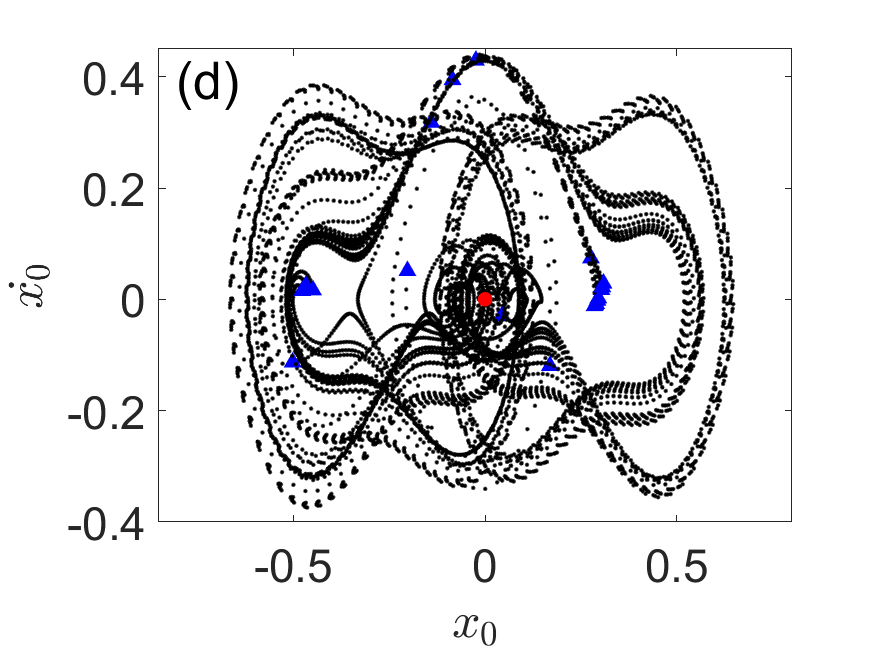}
    \caption{Plots of the phase orbits in the equivalent phase space of ($x_0$, $\dot{x_0}$) from Eq.\;\eqref{con:4}, in the crossover regime as identified in Fig.\;\ref{fig2} of the main text, associated with dynamical instability. (a) $N=11.3\pm0.001$rand(10, 1). (b) $N=13.1\pm0.001$rand(10,1). (c) $N=14.56\pm0.001$rand(10,1). (d) $N=15.29\pm0.001$rand(10,1). The other parameters are $p_1=p_2=0.1, v=0.1$. Here rand(10,1) stands for 10 random numbers uniformly distributed on an interval (0,1). In all plots, the red dots mark the starting points of the trajectories and the blue triangles mark the end points of the $20$ trajectories with different $N$.}
    \label{figS3}
\end{figure}
Here, in order to illustrate  more clearly the chaotic soliton pumping in shallow dynamical superlattices (see Fig.\;\ref{fig2} of the main text), we present the plot of the orbits in the equivalent phase space ($x_0$, $\dot{x_0}$) in Fig.\;\ref{figS3}, calculated for three periods. In all subplots, the orbit starts at ($x_0=0$, $\dot{x_0}=0$) (marked by the red dots), and each subplots contains $20$ trajectories for $20$ random atomic numbers $N$ uniformly distributed on a specified interval [$N_0-0.001$, $N_0+0.001$] [denoted by $N=N_0\pm0.001$rand(10,1)]. As shown in Fig.\;\ref{figS3} (a), we select $N$ in the crossover regime with $0<\mathrm{min}(\mathcal{F})<1$ (see the main text for its definition), and we can see that the soliton is in the pumping state, and that the trajectories for $N=11.3\pm0.001$rand(10,1) are initially identical but separate over time and end up in different locations (marked with blue triangles). This is an indication that the evolution of the centre of mass is dynamically unstable when the initial wavefunction is slightly perturbed. As shown in Figs.\;\ref{figS3} (b)-(d), if we choose $N$ in the crossover regime with $\mathrm{min}(\mathcal{F})=0$, the evolutional trajectories fall onto different attractors, indicating that the soliton undergoes a transition from a pumping state to a trapping state as it evolves over time. The chaotic attractors, as shown in Figs.\;\ref{figS3} (b) and (d),  indicate that a small perturbation of the initial wavefunction can cause the centre-of-mass trajectory to switch between different attractors, leading to a large shift in the centre-of-mass.
\subsection{Pumping and trapping of solitons in deep dynamical lattices}
\begin{figure}[htbp]
	\centering
\includegraphics[trim=20 0 0 0,clip,width=9cm]{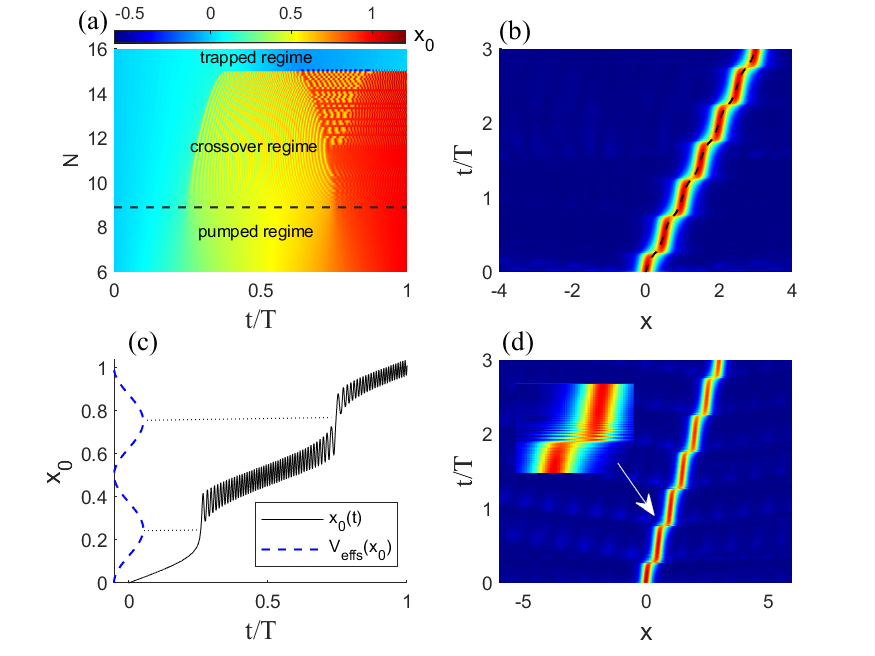}
    \caption{Pumping of bright solitons in deep lattices. (a) Time evolution of the soliton position with different $N$, obtained from the effective equation\;\eqref{con:4} in the main text. Three physically distinct regimes are marked for different ranges of $N$. (b) Spatio-temporal evolution of stable  soliton  pumping for $N = 7$ obtained from the GP equation. The black dashed line visualizes the center-of-mass trajectory predicted by effective-particle dynamics. (c) The one-cycle pumping for the soliton with $N = 10$ predicted by effective-particle dynamics. The dashed line represents the effective potential\;\eqref{con:S5}. (d) Corresponding spatio-temporal evolution of the soliton for $N = 10$. The parameters of the dynamical superlattice are chosen as $p_1=p_2=15,d_1=1/2,d_2=1,v=0.1$.}
    \label{figS4}
\end{figure}

In this section, we give further details about the pumping of the soliton in the deep lattices. As shown in Fig.\;\ref{figS4} (a), the centre-of-mass displacement of the soliton can be identified as three different regimes: a stable pumped regime and a trapped regime, and a crossover regime between the two, accompanied by the emergency of radiations and high-frequency oscillations of the centre-of-mass position. Fig.\;\ref{figS4} (b) shows the centre-of-mass motion of the soliton in the stable pumped regime based on the direct numerical simulation of the GP equation, where the soliton position can be well matched by the prediction of the effective particle dynamics (marked by the black dashed line). As shown in Fig.\;\ref{figS4} (c), when the soliton is in the crossover regime, the numerical results predicted by the effective particle equations indicate that the centre-of-mass displacements undergo leaps near the maximum height of the effective static potential barriers, and each of the leaps is followed by fast oscillations of the centre-of-mass position. The generation of high-frequency oscillations in the centre-of-mass motion is absent in the shallow lattice and can be seen as a competition between strong trapping due to the deep static lattice and pumping due to the sliding lattice. This oscillation is confirmed by the direct numerical simulation of the GP equation as shown in Fig.\;\ref{figS4} (d).
\subsection{Wannier function analysis}
Here we use the same method as~\cite{Fu2022} to analyze the soliton pumping. We expand the soliton state in the basis of the Wannier functions,  $\psi=\sqrt{N}\Sigma_{\alpha,n}a_{\alpha,n}w_{\alpha,n}$, with expansion coefficients $a_{\alpha,n}$, and Wannier functions $w_{\alpha,n}$ labelled by the band index $\alpha$ and the lattice vector  index $n$. The time-dependent coefficients satisfy the normalization condition, $\Sigma_{\alpha,n}|a_{\alpha,n}|^2=1$, and $\rho_{\alpha}(t)=\Sigma_{n}|a_{\alpha,n}|^2$ is the population of the $\alpha$th band. Introducing the full Bloch function $|\varphi_{\alpha k}\rangle=\exp^{\mathrm{i}kx}|u_{\alpha k}\rangle$ in the $\alpha$th band, the Wannier functions can be given by $w_{\alpha,n}=L/{2\pi}\int_{BZ} \varphi_{\alpha k}(x,t)\exp^{-\mathrm{i}knL} dk$, with the quasi-momentum $k \in(\frac{-\pi}{L},\frac{\pi}{L}]$. The Chern number corresponding to the $\alpha$th band can be defined as
\begin{equation}\label{con:S6}
C_{\alpha}=\frac{1}{2\pi\mathrm{i}}\int_{BZ} dk\int_{0}^{T}F_\alpha(k,t)dt,
\end{equation}
where $F_\alpha=\langle\partial_t u_{\alpha k}|\partial_k u_{\alpha k}\rangle-\langle\partial_k u_{\alpha k}|\partial_t u_{\alpha k}\rangle$ is the Berry curvature.
\begin{figure}[htbp]
	\centering
\includegraphics[trim=25 0 0 0,clip,width=8.5cm]{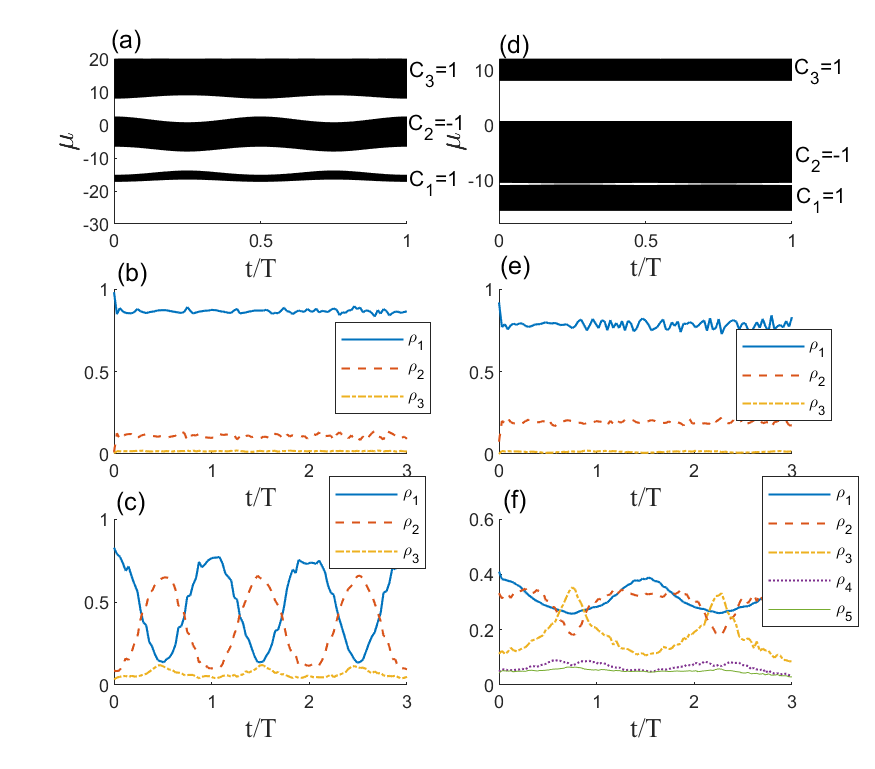}
    \caption{Analysis of linear band occupation for pumped  and trapped solitons (as shown in Fig.\;\ref{fig3} of the main text) in deep lattices, for (a)-(c) (left column) $d_1=1/2$, $d_2=1$ and (d)-(f) (right column) $d_1=1/2$, $d_2=1/3$. (a) and (d): The one-cycle evolution of the first three bands and their Chern numbers. The middle and bottom rows show the evolution of the population of linear energy bands corresponding to (b) and (e) $N=7$, (c) $N=20$ and (f) $N=30$. The moving velocity of the sliding lattice is fixed as $v=0.1$ as before.}
    \label{figS5}
\end{figure}

To illustrate the link between band topology and soliton pumping, we consider lattices where the two lowest bands have Chern numbers with opposite values: $C_1=- C_2=1$. As shown in Fig.\;\ref{figS5}, for relatively low nonlinearity the soliton mainly populates the first band (blue lines) and the population varies slightly with time, while with increasing nonlinearity (increasing $N$) contributions from the upper bands are involved. Fig.\;\ref{figS5} corresponds to Fig.\;\ref{fig3} in the main text. As shown in Figs.\;\ref{figS5} (a)-(c), when $d_1<d_2$, the centre-of-mass displacement [defined as $\langle x(t)\rangle=N^{-1}\int_{-\infty}^{\infty}x|\Psi|^2dx$] of the soliton after each cycle can be roughly estimated by $\langle x(T)\rangle=\Sigma_{\alpha}\rho_{\alpha}C_{\alpha}L_0$, which is determined by the averaged contributions of the Chern numbers of the occupied energy band~\cite{Fu2022}. In this case, as shown in Fig.\;\ref{figS5} (c), the Rabi oscillations between the Bloch bands carrying topological transport in opposite directions can lead to the soliton trapping for a strong nonlinearity [see the Fig.\;\ref{fig3} (b) of the main text]. However, it is noteworthy that when $d_1>d_2$, as shown in Fig.\;\ref{figS5} (e), for relatively low nonlinearity, the soliton predominantly occupies the first energy band, but the soliton is trapped [see Fig.\;\ref{fig3} (c) of the main text], which does not correspond to the first Chern number of the pump. This suggests that using the perspective from time-dependent band occupation  and the Chern number of the linear energy band does not fully reflect the pumping and trapping of the soliton.
\subsection{Calculation of gap solitons}
\begin{figure}[htbp]
\includegraphics[trim=20 0 0 0,clip,width=8cm]{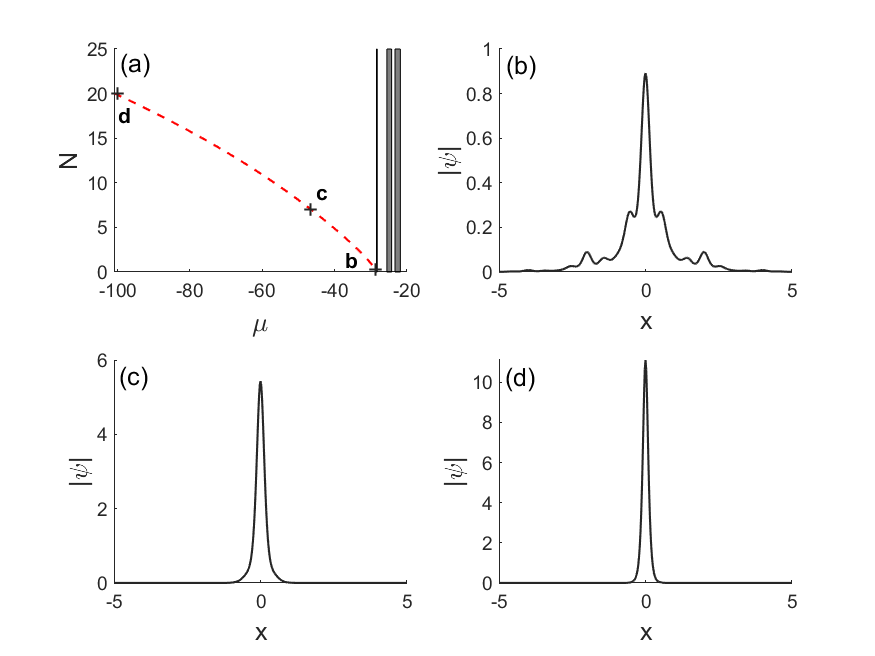}
    \caption{(a) The norm $N$ as a function of the chemical potential $\mu$ for the gap soliton in a semi-infinite bandgap bifurcating from the first lowest band. The shaded areas are linear bands. (b) , (c) and (d) are the profiles of gap solitons with $N=0.3, 7$ and $20$, corresponding to points b, c, and d labelled in panel (a). The lattice parameters are $p_1=p_2=25, d_1=\frac{1}{2}, d_2=\frac{2}{3}$.}
    \label{figS6}
\end{figure}
It is well known that nonlinear periodic systems allow the existence of spatially localized eigenstates, the so-called gap solitons, which can be found numerically. We use the square operator method~\cite{Yang2007} to compute the gap soliton bifurcating from the lowest band in a semi-infinite bandgap, using $\mathrm{max}(|\psi_{n+1} - \psi_{n}|)<10^{-10}$ as a convergence criterion at the end of the iteration to ensure that the gap soliton achieves the desired accuracy, where $\psi_{n+1}$ and $\psi_{n}$ are the wavefunction of calculated soliton for the $n$th iteration and $n+1$th iteration, respectively. In Fig.\;\ref{figS6} (a), we have plotted the $N$ as a function of the chemical potential $\mu$, for the gap solitons bifurcating from the lowest linear energy band with increasing $N$. Soliton profiles corresponding to the labelled points in Fig.\;\ref{figS6} (a) are shown in Figs.\;\ref{figS6} (b), (c) and (d) respectively. As $N$ increases, the gap solitons in the semi-infinite bandgap become more spatially localized and the particle nature of the gap solitons becomes more prominent, which is consistent with the nature of the trial wavefunction of the bright soliton for which the effective equation\;\eqref{con:4} was obtained in the main text. These numerically calculated nonlinear eigenstates (i.e. the gap solitons with different $N$ as shown in Fig.\;\ref{figS6}) of the instantaneous nonlinear Hamiltonian are used as initial excitations to explore the pumping and trapping of gap solitons in adiabatic time-varying superlattice systems in the main text.

\begin{figure*}[htbp]
\includegraphics[trim=0 0 0 0,clip,width=17cm]{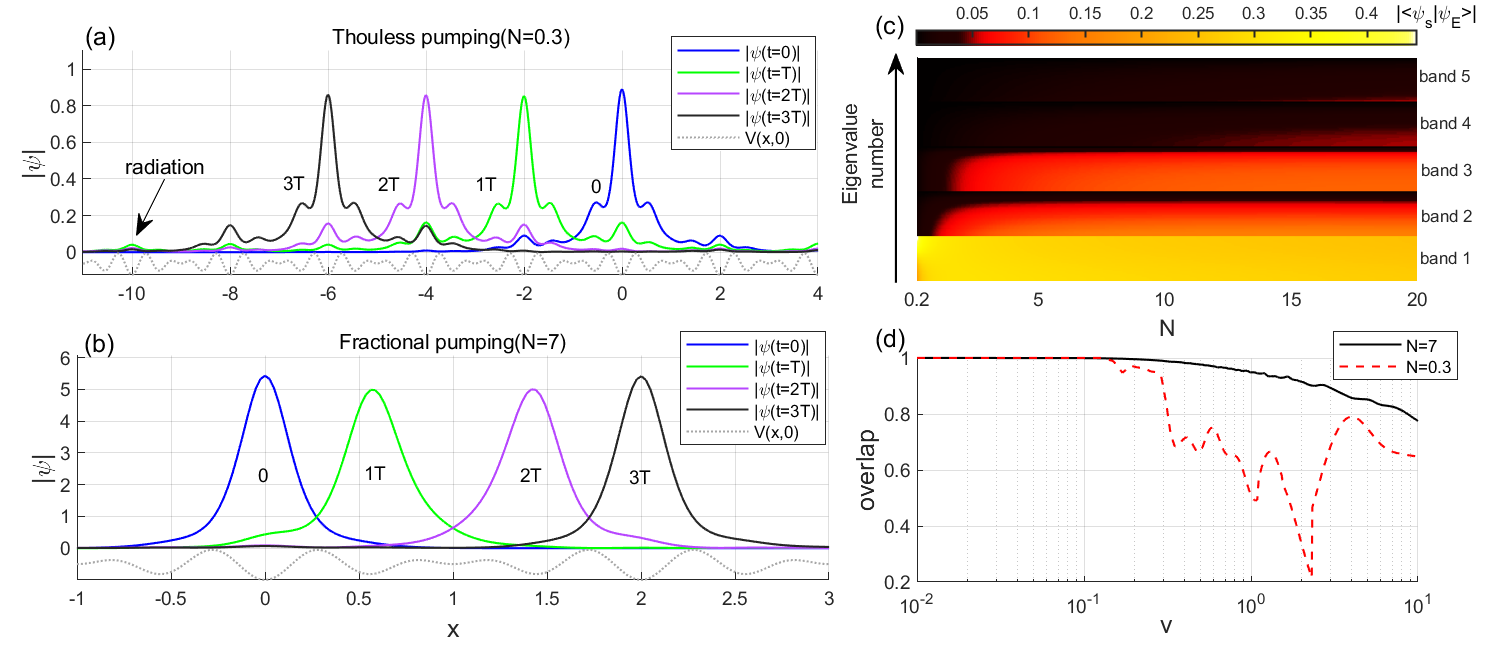}
    \caption{The shapes of the propagated solitons at $t=0$ and after each full period for (a) $N=0.3$ and (b) $N=7$. The excitation is an
instantaneous gap soliton. The lines at the bottom represent the periodic potential at $t=0$, which returns to its initial form at the end of each cycle with an adiabatic velocity of $v=0.1$. (c) Projection of the initial soliton wavefunctions onto linear energy eigenstates. The linear eigenstates in each band are arranged in ascending order of their eigenvalues from bottom to top. (d) Overlap of a numerically propagated soliton with the instantaneous soliton after one pump cycle, for increasing the moving lattice velocity $v$. The lattice parameters are $p_1=p_2=25, d_1=\frac{1}{2}, d_2=\frac{2}{3}$.}
    \label{figS7}
\end{figure*}
\subsection{Pumping and trapping of gap solitons in deep dynamical lattices}
In this last section, we provide some further details about Thouless pumping and fractional pumping of gap solitons in adiabatic evolutions. Figs.\;\ref{figS7} (a) and (b) show the numerically propagated wavefunctions of solitons at $t = 0, T, 2T$, and $3T$ for different nonlinearities $N = 0.3$ and $N = 7$, respectively. As shown in Fig.\;\ref{figS7} (a), for a weak nonlinear soliton ($N=0.3$) in a Thouless pump, radiation is generated at the beginning of the pumping, and then the soliton returns to the same wavefunction after each pumping cycle, shifted by only one unit cell, as dictated by the Chern number of the band from which the soliton bifurcates. The adiabatic propagation of a soliton for stronger nonlinearity ($N=7$), showing fractional pumping, is shown in Fig.\;\ref{figS7} (b). In this case, after one period, the soliton is clearly different from the soliton at $t = 0$, and the wavefunction is localized to a neighbouring potential minimum within a unit cell, with the centre-of-mass position displaced by $1/3$ unit cell, and only after three periods the soliton returns to its initial state, apart from a translation by one unit cell. We numerically confirm that the fractionally pumped solitons at $t=T, 2T$ are new instantaneous gap solitons generated by nonlinear bifurcation, localized at different potential minima within a unit cell, with the same norm $N$ as the instantaneous gap soliton at $t=0$, but with a different chemical potential $\mu$. This bifurcation is unique to nonlinear systems, where the linear potential returns to its initial form after each period, whereas the nonlinear potential, which depends on the nonlinear eigenstate (gap soliton), may not return to its initial state, creating new gap solitons. During the adiabatic evolution, the soliton then follows these eigenstates, and when the nonlinear periodic Hamiltonian (which consists of the linear time-varying potential and the squared modulus of  nonlinear eigenstates) comes back to itself-apart from a translation by an integer number of unit cells due to the translation invariance-after multiple periods, the soliton is forced to transport by an integer number of unit cells. In Fig.\;\ref{figS7} (c) we also examine the projection of the soliton wavefunction onto the linear energy eigenstates. At low nonlinearity (e.g. $N=0.3$), the soliton occupies the lowest band from which it bifurcates. As the nonlinearity increases, the soliton starts to occupy energy eigenstates from the higher bands, but the occupation is never perfectly uniform. For example, at $N=7$, which corresponds to Fig.\;\ref{fig4} (c) in the main text, the soliton occupies the three lowest bands. Fig.\;\ref{figS7} (d) confirms that the evolution of the stable solitons can be well described by instantaneous solitons in the adiabatic limit. We numerically calculated the overlap, $\left|\int_{-\infty}^{\infty}dx\psi^{*}(T)\Psi(T)\right|/N$, between the instantaneous soliton $\psi(T)$ and the propagated soliton $\Psi(T)$ after one period by initializing the system with an instantaneous soliton for $t = 0$. The perfect overlap at sufficiently low moving sublattice velocity confirms the existence of an adiabatic limit for soliton pumping. During the adiabatic evolution, due to the slow variation of the superlattice potential with time, the soliton behaves like a classical particle, moving to the next local potential  minimum after each cycle, as the pumping proceeds.

\begin{figure}[htbp]
\includegraphics[trim=20 0 0 0,clip,width=8.2cm]{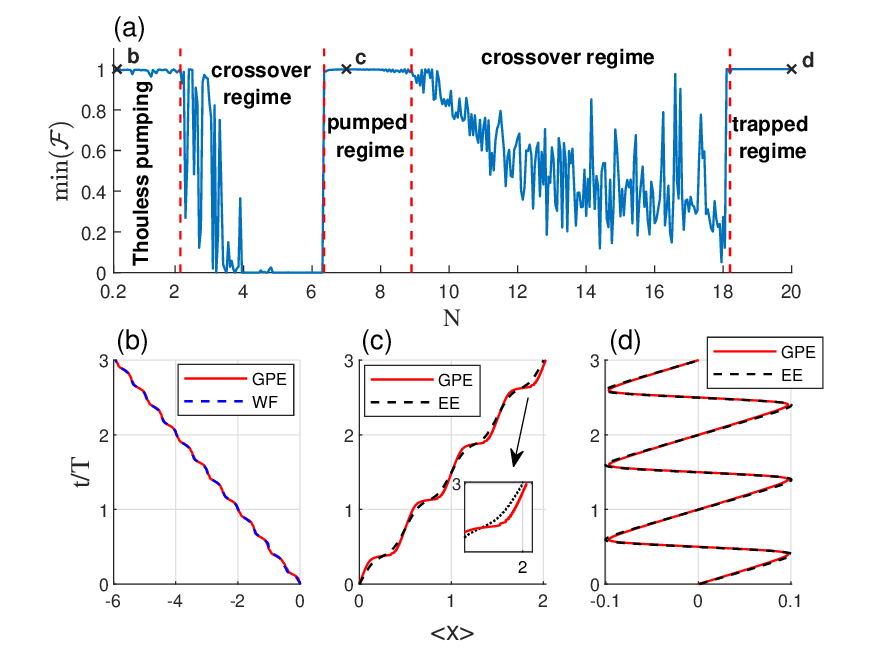}
    \caption{(a) Minimum of the quantum fidelity [see Eq.\;\;\eqref{con:5} in the main text] attained during the evolution of the soliton as a function of $N$. Physically distinct regimes are identified and labelled. (b) Integer Thouless pumping for the gap soliton with $N=0.3$. A comparison is made between the centre-of-mass trajectory of the numerically propagated soliton and that of the relevant instantaneous Wannier function (WF). (c) Fractional pumping for the gap soliton with $N=7$. The one-cycle displacement of the gap soliton is $L/3$ (where $L_0=2$). (d) Trapping of the gap soliton with $N=20$. The lattice parameters are $p_1=p_2=25, v=0.1, d_1=\frac{1}{2}, d_2=\frac{2}{3}$. In panels (c) and (d), the changes in the soliton position are given by both the GP equation (GPE) and the effective equation (EE) with the effective-particle approximation. In all cases, the excitation is an instantaneous gap soliton.}
    \label{figS8}
\end{figure}

 In Fig.\;\ref{figS8} (a) we calculate the minimum value of quantum fidelity attained  during the time evolution of the gap solitons and also find that there exist physically distinct regimes: the stable pumped (Thouless pumping, fractional pumping) regime, the trapped regime and the crossover regime associated with the dynamical instability. Three typical time-evolution behaviours for the centre-of-mass displacement of the gap solitons are shown in Figs.\;\ref{figS8} (b), (c) and (d), respectively. Fig.\;\ref{figS8} (b) shows the integer Thouless pumping of solitons, where the gap soliton is stable and follows the position of the instantaneous maximally localized Wannier function (WF) for the first lowest band. It should be noted that if the constraint $\frac{p_j}{2N^2}(j=1,2)$ is not satisfied, $\tilde{V}(X,\tau)$ [see the right-hand term of Eq.\;\eqref{con:6} in the main text] cannot be considered as a perturbation. In this case, the centre-of-mass trajectory for $N = 0.3$ shown in Fig.\;\ref{figS8} (b) cannot be governed by the effective-particle dynamics because the effective particle approximation is no longer valid for sufficiently small $N$. Figs.\;\ref{figS8} (c) and (d) show the fractional pumping and trapping of the gap solitons, respectively, where the center-of-mass motion is well captured by the effective-particle dynamics using the effective equation (EE)\;\eqref{con:4} in the main text. Although the shape of the fractionally pumped soliton changes with time, probably because the soliton shape changes very slowly, the centre-of-mass motion of the soliton can still be accurately described in terms of the effective-particle dynamics. We note that the fractional displacement of the pumped gap soliton can also be explained very nicely in terms of multi-band Wannier functions, as proposed in Ref.~\cite{M2023}, i.e. the solitons follow the centre-of-mass positions of the respective maximally localized multi-band Wannier functions. We also note that the fractional displacement (in units of the lattice vector $L$) per period can be accounted for by the formula $f=\Sigma^{N_b}_{i=1}C_i/N_b$ introduced in Ref.~\cite{M2023}, where $N_b$ denotes the number of bands involved and the numerator defines the sum of the Chern numbers $C_i$ of the respective bands. For the case shown in Fig.\;\ref{figS8}(c) (corresponding to Fig.\;\ref{fig4}(c) in the main text), $N_b=3$, and thus $f = (-1+3-1)/3 = 1/3$.

\begin{figure}[htbp]
\includegraphics[trim=20 0 0 0,clip,width=9cm]{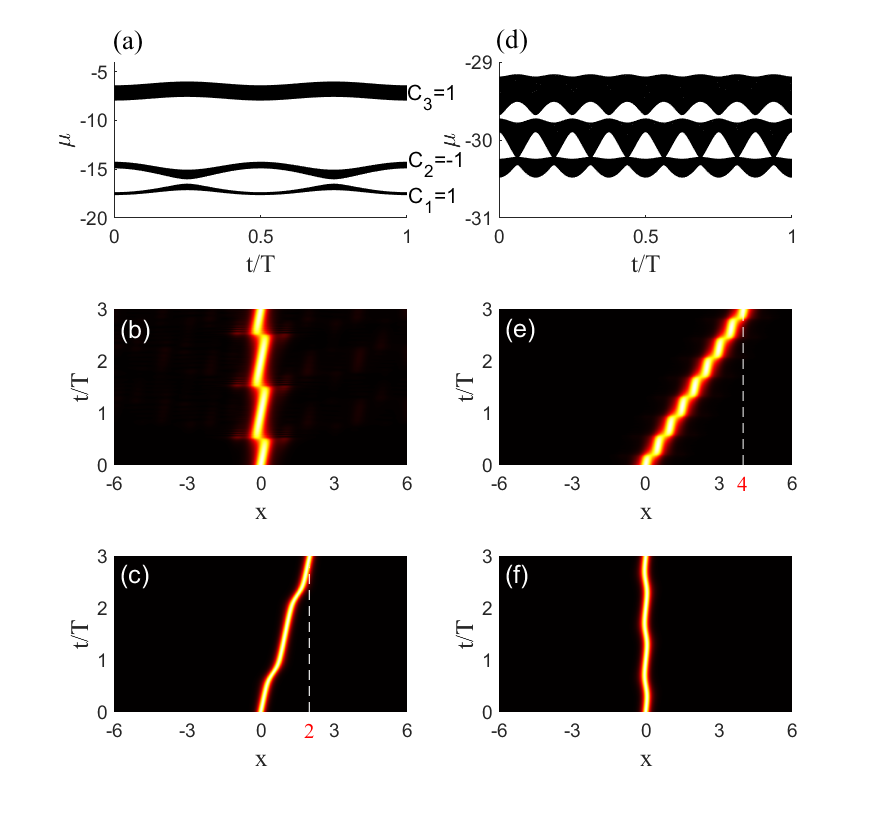}
    \caption{The linear band structure showing the lowest three bands and the spatio-temporal evolution of the gap solitons for the cases $p_1=p_2=15$, $d_1=1$, $d_2=\frac{2}{3}$ [(a), (b) and (c)] and $p_1=p_2=25$, $d_1=\frac{1}{2}$, $d_2=\frac{4}{3}$ [(d), (e) and (f)]. (c) and (e): pumped solitons. (b) and (f): trapped solitons. The excitation is an instantaneous gap soliton of the system with different degrees of nonlinearity (normalized particle number): (b) and (e) $N =4$, (c) and (f) $N =20$. The other parameters are $v=0.1$, $T=10\pi$.}
    \label{figS9}
\end{figure}

In Fig.\;\ref{figS9} we consider the evolution of the gap solitons for the cases $d_1>d_2$ (the left column) and $d_1<d_2$ (the right column), with the one-cycle evolution of the lowest three bands shown in Figs.\;\ref{figS9} (a) and (d) respectively. As shown in Figs.\;\ref{figS9} (a)-(c), for $d_1>d_2$, in the intermediate-nonlinearity regime ($N=4$ in our simulation) the gap soliton is trapped, while for stronger nonlinearity (corresponding to $N=20$) the pumping of the gap soliton occurs. In contrast, for $d_1<d_2$, the increase in nonlinearity (increasing $N=4$ to $20$) instead leads to a transition from a pumped soliton to a trapped soliton, as shown in Figs.\;\ref{figS9} (e) and (f). In both cases, the one-cycle-averaged displacements of the pumped gap solitons are approximately $d_2$. In Fig.\;\ref{figS9} (d), we note that the first linear band gap closes, and thus there is no well-defined Chern number, which  demonstrates that the bandgap closings within the group of participating bands do not change the transition between  fractional pumping and trapping of the solitons. So far, all the numerical observations of the evolution of the gap solitons are fully consistent with the predictions obtained from the effective Eq.\;\eqref{con:4} in the main text, which again suggests that the particle nature of the solitons plays a dominant role in the transport.
\end{document}